\documentclass[12pt,twocolumn]{article}
\usepackage[a4paper,left=20mm,right=20mm,top=25mm,bottom=25mm,includeheadfoot]{geometry}

\usepackage{mathpazo}
\usepackage{helvet}
\usepackage{graphicx}
\usepackage{amsmath,textcomp}
\usepackage{makeidx} 
\makeindex
\usepackage{sectsty}
	\allsectionsfont{\sffamily\raggedright}
	\sectionfont{\sffamily\large\raggedright}
	\subsectionfont{\sffamily\normalsize\raggedright}
\usepackage{ftnright}

\usepackage{widetext}
\usepackage{flushend}
\usepackage{cuted}
\usepackage{color}
\usepackage{hyperref}
\usepackage{pstricks,amssymb}

\usepackage{fancyhdr}
\begin{document}
%Do not change the \vspace command, which is needed to suppress extra space above the title.
\title{\vspace{-2em}\bfseries\sffamily Is interstellar travel to an exoplanet possible?}
\author{\normalsize Tanmay Singal${^1}$ and 
Ashok K. Singal${^2}$\\[2ex]
$^{1}$Department of Physics and Center for Field Theory and Particle Physics,\\ Fudan University, Shanghai 200433, China.\\
{\tt tanmaysingal@gmail.com}\\[2ex]
$^{2}$Astronomy and Astrophysis Division, Physical Research Laboratory\\
Navrangpura, Ahmedabad 380 009, India.\\
{\tt ashokkumar.singal@gmail.com}
}
\date{\itshape Submitted on xx-xxx-xxxx}
\maketitle
\thispagestyle{fancy}

%Do not change the \sffamily command, it is needed to ensure that the abstract appears in a different font.
\begin{abstract}
{\sffamily
In this article, we examine the possibility of interstellar travel to reach some exoplanet orbiting around a star, beyond our Solar system. Such travels have been in the realm of science fiction for long. However, in the last 50 years or so, this question has gained further impetus  in the mind of a man on the street, after the interplanetary travel has become a reality. Of course the distances to be covered to reach even some of the nearest stars outside the Solar system could be hundreds of thousands of time larger than those encountered within the interplanetary space. Consequently, the time and energy requirements for such a travel could be immensely prohibitive. The questions we want to explore here are: What could be the possible limitations, if any, for such interstellar travels, and could humans ever undertake such a  voyage, with hopefully a positive outcome? What could be a possible scenario for such an adventure in a near or even distant future? And what could be the reality of UFOs -- Unidentified Flying Objects -- that get reported in the media from time to time?
}\\ 
\hrule
%Do not change the \hrule command, it is needed to separate the abstract form the main text.
\end{abstract}
%--------------------------------------------------
\section{Introduction}
In the last three decades many thousands of exoplanets, planets that orbit around stars beyond our Solar system, have been discovered. Many of them are in the habitable zone, possibly with some forms of life evolved on a fraction of them, and hopefully, the existence of intelligent life on some of them. Can we ever get in physical contact with the extra-terrestrials, assuming they are there? Radio communication over interstellar distances is one possibility \cite{5}. What about the possibility of humans ever visiting ``them''? Or an even more pertinent question first -- Is an interstellar space travel to an exoplanet around a star beyond our Solar system possible? 

In last 50-60 years, the mankind, first time in its history, has not only ventured into outer space, humans have successfully stepped on the Moon, the first time ever on another celestial body. Rover explorations of the surface of Mars have been made many times, probes have landed on  Venus, and many other missions have been sent to other planets. The Galileo spacecraft that entered orbit around Jupiter, made a number of close flybys to study Jupiter's satellite Ganymede. In the Cassini-Huygens mission, while Cassini orbited Saturn and studies its rings before it plunged into Saturn’s atmosphere, the Huygens probe successfully landed on Saturn's moon Titan.

In recent years, India too has sent two missions, Chandrayan-1 and Chandrayan-2, to the Moon, and Mars Orbiter Mission (MOM), India's first interplanetary mission, has successfully reached Mars. A third mission to the Moon is now being planned, and other interplanetary missions are in the offing. Perhaps in a decade or so, India may also achieve a human landing on the the Moon. After that one could imagine such manned trips to Mars. Other countries are also planning such expeditions in near future. As for the Jovian planets like Jupiter or Saturn, manned missions if any, will have to have bases on one of their satellites, e.g. Ganymede or Titan, as the planets themselves are all gaseous, lacking a solid surface to make a landing. 

This begs a question: Could man possibly ever travel to distant stars to visit some exoplanets, perhaps in a habitable zone, to possibly encounter some extraterrestrial life? After all, a mere  century back, a trip to the Moon, culminating in a human landing on it, looked as much impossible  and such accounts in science fiction seemed to be just a fig of imagination, as an interstellar travel to an exoplanet may appear now. Such analogies though may have their own justification grounds, but the fact remains that the distances involved in interstellar travel are immensely larger. 
The nearest star outside the solar system (Proxima Centauri) is as many times ($\sim$ a hundred million times) farther than the Moon, as the latter is compared to distance between adjacent rooms (($\sim 4$ m) in a building. 
From a simple logic one could then expect that going to a star will at least be as much more difficult than going to the Moon as the going-to-the-Moon was with respect to a walk just next door within an office building. Of course, the shortness of human lifetime makes things all the more difficult. 
With the maximum speeds achieved so far by the spaceships within the solar system, it will require about 80,000 years on a one-way journey to this nearest star. Thus it may not look possible to reach other stars within a human lifetime, although on a theoretical basis theory of relativity could allow one to do so. For instance, a spaceship accelerating continuously with a convenient value of $g$, that is the acceleration that we are used to on the surface of the Earth, 
could travel to the most distant parts of the universe within a human lifetime, 
without violating the speed-limit of $c$, the speed of light. 
In principle, interstellar travel may thus appear possible.

However, energies involved in such an endeavour would make it next to impossible. In a spaceship the fuel needed for the later parts of the journey has to be carried aboard and thus also needs to be accelerated till it is utilized. Therefore the initial mass at the start of such a voyage is exponentially larger than the final payload.  With conventional chemical fuel such an arduous journey will need a fuel-mass of a whole galaxy. Even within the best possible scenario, where almost $100\%$ of mass is converted into energy (in a typical thermonuclear reaction only about $0.7\%$ of mass is converted into energy), one would require initial mass to be millions of times the mass of the final payload and the energy required may be worth hundreds of years of total energy consumption of the whole world. If we imagine that the energy is beamed from power plants on the Earth to the spaceship, it will again require many hundred million megawatts of power throughout the duration of such a trip, which might last for a very long time. It therefore looks that at most we might travel to other planets within our solar system but the distant stars will ever remain within the realm of a distant dream only.

In this article, we ignore the technical aspects of the mission as technology is bound to improve rapidly over time. Further, we assume 100\% efficiency of the rocket engine in converting fuel energy into kinetic energy of the exhaust, something that might not really be possible. We carry forth the possibility of our endeavour without delving into many other equally important issues such as the long term effects of cosmic radiation on the health of space travellers and their requirements for food, medical and other life-sustaining needs. We consider mainly the minimum basics of the travel, which are distance, time and energy.

\section{The story so far}

Till date there have been five spacecrafts that have crossed the threshold of escape velocity from the solar system and four of them are already headed towards the interstellar space.

Pioneer 10 was launched in 1972, flew past Jupiter in 1973 and became the first spacecraft to achieve escape velocity from the solar system. The contact was lost in January 2003 and is heading in the direction of Aldebaran in Taurus. Pioneer 11 was launched in 1973, flew past Jupiter in 1974 and Saturn in 1979. The contact was lost in November 1995. The spacecraft is headed toward the constellation of Aquila.

Pioneer 10, as well as Pioneer 11, carry gold-anodized aluminium plaques in case either spacecraft is ever found by intelligent life-forms from another planetary system. The plaques feature the human figures along with several coded-symbols that are designed to provide information about the origin of the spacecraft, and the message may hopefully survive for hundreds of millions of years during its long travel through the interstellar space. It is, thus, the artefact of mankind with the longest expected lifetime \cite{2}.
%--------------------------
\begin{figure}[t]
\includegraphics[width=\linewidth]{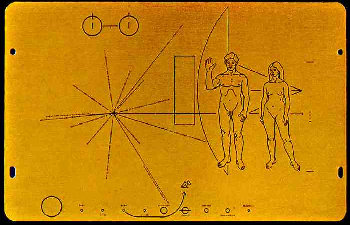}
\caption{The message, featuring human figures along with several coded-symbols inscribed on the gold-anodized aluminium plaques, carried aboard Pioneer and Voyager spacecrafts.}
\end{figure}

The content of the message should be clear to an advanced extraterrestrial civilization, which will have, of course, the entire Pioneer 10 spacecraft itself at its disposal to examine as well. But being the product of billions of years of independent biological evolution, they may not at all resemble humans, nor may the perspective and line-drawing conventions be the same there as here. The human beings will perhaps be the most mysterious part of the whole message for them \cite{2}.

Voyager 1 was launched in September 1977, flew past Jupiter and Saturn, made a  close approach to Saturn's moon Titan and is now at a distance of about 145 astronomical unit (au), where one au (=$1.5 \times 10^8$ km) is the average distance of the Earth from the Sun. Voyager 2 was launched in August 1977, flew past Jupiter, Saturn, Uranus, Neptune and is now at a distance of about 125 au. Both probes are already past heliopause, the region where the solar wind interacts with the interstellar medium at distances around 120 au from the Sun. Voyagers are thus presently exploring the boundary between the Sun's influence and interstellar space, where nothing from the Earth has flown before, and are expected to return valuable data, hopefully, for another decade. Since the Pioneers were launched first, they had a head start on the Voyagers, but because they were travelling slower they were eventually overtaken by Voyagers. 

New Horizons, launched in 2006, made a flyby of Jupiter in 2007, and then in 2015 it made a flyby of Pluto, where it flew 12,500 km  above the surface of Pluto, making it the first spacecraft to explore this dwarf planet. After  that, New Horizons made a flyby of Kuiper belt object 486958 Arrokoth, at $\sim 43$ au from the Sun. New Horizons was launched with the largest-ever launch speed for a man-made object. It will, however, slow down to an escape velocity of only 2.5 au per year as it moves away from the Sun, and it will never overtake the Voyagers.
\subsection{The Pale Blue Dot}
The pale blue dot is a photograph of planet Earth taken in 1990 by the Voyager 1 spacecraft 
when the spacecraft reached $\sim 6$ billion km, or about 40 au (the distance of Pluto), from the Sun. 
This is an actual photograph (Fig. 2) of the Earth, taken from the farthest distance till now,  
and it appears as a tiny pale blue dot against the background of an apparent void  
(the faint brown band is due to the reflection of sunlight from camera optics). 
This picture is very significant as a perspective on our place in the cosmos as our blue planet literally pales into insignificance within the larger scheme of things. And this is the only actual image of the Earth ever seen by anybody from such a vantage point. It is both a chastening and humbling realization for us humans that our huge planet is such a tiny speck of dust seen from the distance of an outpost (Pluto!) of our planetary system. If it could be photographed from near our nearest star [Proxima Centauri], its diameter will appear about 7000 times smaller and it would be still fainter in brilliance by a factor of 50 million (with the flux-density falling as a square of distance), and that the Earth may not even qualify to be called ``a tiny speck of dust'' from our just next-door neighbour star.
%--------------------------
\begin{figure}[t]
\includegraphics[width=\linewidth]{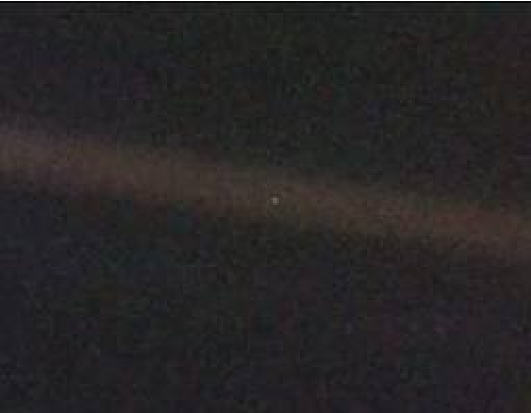}
\caption{A panoramic (!) view of our Earth, that appears as a pale bluish dot in the centre of the image. The faint brown band across the image is due to the reflection of sunlight from camera optics.}
\end{figure}

As Carl Sagan writes \cite{3} ``The Earth is a very small stage in a vast cosmic arena. Think of the rivers of blood spilled by all those generals and emperors so that in glory and triumph they could become the momentary masters of a fraction of a dot. Think of the endless cruelties visited by the inhabitants of one corner of this pixel on the scarcely distinguishable inhabitants of some other corner. How frequent their misunderstandings, how eager they are to kill one another, how fervent their hatreds. Our posturings, our imagined self-importance, the delusion that we have some privileged position in the universe, are challenged by this point of pale light. Our planet is a lonely speck in the great enveloping cosmic dark. In our obscurity - in all this vastness - there is no hint that help will come from elsewhere to save us from ourselves. The Earth is the only world known, so far, to harbour life.''

\begin{table*}[t]
\begin{center}
\caption{An idea of the cosmic distances involved}
\vspace{-0.3cm}
\vbox{\columnwidth=33pc}
\begin{tabular}{|c|c|}
\hline
{\bf Cosmic object}	  & {\bf Distance}\\ 
\hline
Moon&	 1.28 light seconds	(384000 km)\\
\hline
Sun	& 500 light seconds (150 million km)	\\
\hline
Proxima Centauri	& 4.24 light years	\\
\hline
Orion Nebula&	1300 light years	\\\hline
Centre of  Milky-way&	25,000 light years\\	\hline
Andromeda Galaxy	& 2 million light years	\\\hline
Size of Universe! &	14 billion light years	\\
\hline
\end{tabular}
\end{center}
\end{table*}

And to think further that somewhere on a far-off world perhaps some intelligent being looking at this ``not-even-a-speck-of-dust'' could amusedly imagine that some two-legged creatures, populating that utterly insignificant part of the universe, believe that some of their ancestors (saints, gurus or prophets), confined to a minuscule part of this tiniest of dots, had figured out the grandest design of the whole Universe or even of its so-called creator -- and have the audacity to claim that the creator himself or his some messenger had appeared in the form of these very two-legged creatures on their own planet. It should also humble us and put into total insignificance the occurrence of all our daily squabbles, aspirations, the desire to preserve our DNA through our children and grandchildren, political upheavals, love-affairs, wars between nations, and above all it should show us the hollowness of our religious beliefs -- perhaps the greatest folly of all -- and our chauvinism that we are the best of all, with an utter contempt for others who may not agree with us, and our willingness to condemn those others or even kill and die for some totally unfounded beliefs uttered or penned down by someone perhaps with good intentions but based on the limited knowledge at that moment of time, or much worse, based on a  pure whim and fancy, thrust upon other gullible fellow beings.

\section{Cosmic distances involved}
The main challenge facing interstellar travel is the vast distances that have to be covered, requiring very high speeds as well as long travel times. The latter make it particularly difficult to design manned missions. 

\subsection{How far can a manned mission travel from Earth?}
 
As one cannot travel faster than light, one might conclude that a human can never make a round-trip farther than 20 light years (1 light year $\approx 9.5 \times 10^{17}$, the distance travelled by light in one year), assuming the traveller is active between the ages of 20 and 60. 
Thus one would never be able to go beyond a few star systems which exist within the limit of $\sim 20$ light years from the Earth. Even if we design a spaceship that can travel at $0.99c$, where $c\approx 3 \times 10^{10}$ cm/sec is the speed of light, which, from the theory of relativity is the maximum possible speed an object could ever attain, interstellar travel beyond some  nearest stars seems impossible.

To survive for long years on a spaceship, it would be ideal to maintain a constant acceleration,  $g \approx 9.8 \times 10^2$ cm sec$^{-2}$, the acceleration due to gravity that the humans have evolved in and are accustomed to on the Earth, with the rocket continuously accelerating the spaceship by this amount. 
Since we may want soft landings on the surface of the exoplanet as well as on our return to the Earth, we divide our journey into four separate stages. In the onward Journey while the spaceship is moving towards the destination, it will be accelerated in the first half of the journey, while in the second half it will have to be decelerated to attain an almost zero speed. In the same way during the return journey, it will have to be accelerated in first half of the return journey and then decelerated in the second half, for an ultimate soft landing.
\begin{table*}[t]
\begin{center}
\caption{Effects of time dilation}
\vspace{-0.3cm}
\vbox{\columnwidth=33pc}
\begin{tabular}{|c|c|c|}
\hline
{\bf Time on spaceship}&{\bf Time on Earth} & {\bf Distance reached}\\
$T$ (years)& $t$ (years)& $d$ (light years)\\
\hline
1& 1.01 &0.065\\\hline
2 &2.1& 0.26\\\hline
5& 6.5 &1.85\\\hline
7 &11.5 &4.1\\\hline
10 &25 &11\\\hline
15& 93 &45\\\hline
20& 340 &165\\\hline
25& 1,230& 610\\\hline
30& 4,450 &2,200\\\hline
40& 59,000 &29,000\\\hline
50& 780,000 &390,000\\\hline
60& 10,100,000 &5,100,000\\\hline
90& 23,500,000,000 &11,700,000,000\\
\hline
\end{tabular}
\end{center}
\end{table*}

\subsection{The relativity comes to the rescue -- time dilation}
A constant acceleration of $1g$ for a year would bring the speed of spaceship approximately close to $c$. Therefore relativistic effects of time dilation would have to be taken into consideration. We know that time passes relatively slower by a relativistic factor 
 $\gamma= 1/\sqrt{1-(v/c)^2}$ for an observer moving with a relative speed $v$.
Detailed calculations show that by the time the spaceship lands back on the Earth, the time 
$t$, that would have passed on the Earth, 
would be related to the time $T$, that passed on the spaceship, as (see Appendix A)
\begin{equation}
t = \frac{4c}{g} \sinh \frac{gT} {4c},
\end{equation}
a factor of 4 in the formula appears because of the four stages of the journey.  During this time, the maximum relative speed the spaceship would achieve, midway of the journey, is
\begin{eqnarray}
{v}= {c}\tanh (gT/c)
\end{eqnarray}
The maximum distance $d$, of the destination that the spaceship would have arrived at and returned from, will be given by 
\begin{equation}
d = \frac{2c^2}{g} \left[\cosh \frac{gT}{4c}-1\right].
\end{equation}
The destination distance $D$ can be expressed in terms of time $t$ of the Earth, as
\begin{eqnarray}
D = \frac{2c^2}{g} \left[\sqrt{1+\left(\frac{gt}{4c}\right)^2}-1\right].
\end{eqnarray}
It helps to remember that for $g \approx 9.8 \times 10^2$ cm sec$^{-2}$, time $c/g =0.97 \approx 1$ year and the distance $c^2/g \approx 1$ light year.
Table 2 gives us an idea of the time dilation involved from the total duration and distance reached in a round trip, involving a constant acceleration of $1g$ for the crew. A future spacecraft, using technologies that we haven't even dreamed of, may use an engine that could sustain a constant acceleration of $1g$. 
Travelling even at the speed of light, visiting the stellar nursery in Orion nebula would require at least 2600 years on the earth time, while a cruise to the centre of our Milky-way galaxy will take more than 50,000 years, and a round trip to Andromeda, the nearest spiral galaxy, will need at least 4 million years. But due to the relativistic time dilation, for the traveller the time spent could be much smaller. With a $1g$ engine, a vacation trip to Andromeda may be possible within a human lifetime! 

For those astronauts, however, returning back home is out of the question. Back on the Earth, millions of years would have passed and entire civilizations would have come and gone, while the astronauts who left in their twenties would be still in their eighties.
Table 2 gives time spent by the astronaut, travelling in a rocket with a constant acceleration, as a function of time and distance seen from the Earth.

\section{The rocket equation}
If the fuel needed for the journey has to be carried aboard it also needs to be accelerated till it is utilized. Therefore the initial mass, $M_i$, at the start of the journey is much more than $M_f$, the final payload mass. This is given by the rocket equation, which gives the final reachable speed $v$ as a function of the exhaust speed $u$ of gas/ion/light emission and ${\cal R}=M_i/M_f$, the ratio of the initial mass (payload + fuel) to the final mass (only payload). From the momentum conservation we have 
\begin{equation*}
 M \frac{dv}{dt}=-\frac{dM}{dt} u, 
\end{equation*}
We can integrate it
\begin{equation*}
\int_{0}^{v}{{\rm d} v}=-u\int_{M_i}^{M_f}\frac{{\rm d}M}{M} ,
\end{equation*}
 which gives
\begin{equation*}
\frac{v}{u}= \ln {\cal R}. 
\end{equation*}
  or
\begin{equation}
{\cal R}  = \exp(v/u).
\end{equation}
The exponential makes the required mass ratio increase very fast with $v/u$. \\
For example,\\ 
${\cal R}  = 1\:,\:$ for $v = 2.3 u$,\\  but\\
${\cal R}  = 10^{10}\:,\:$ for $v = 23 u$.

Thus, to obtain a final speed, $v$ close to $c$, it is necessary for $u$ to be of the order of $c$ as well, otherwise the required mass ratio will be prohibitively large.

In a relativistic case, the rocket equation becomes (see Appendix B)
\begin{equation*}
\frac{v}{c} = \frac{1 - {\cal R}^{-2u/c}} {1 + {\cal R}^{-2u/c}}.
\end{equation*}
For the mass ratio ${\cal R}$, we then get
\begin{equation*}
{\cal R}  = \left[\frac{(1 + \frac{v}{c})}{(1 - \frac{v}{c})}\right]^{c/2u},
\end{equation*}
or 
\begin{equation}
{\cal R}  = \left[\gamma{\left(1 + \frac{v}{c}\right)}\right]^{c/u}.
\end{equation}
For $v\ll u \le c$, equation (6) reduces to the familiar non-relativistic equation (5).

The power of the rocket engine needed can be calculated from the required thrust of the rocket, which is nothing but the total mass, $M_i$, of the spaceship (payload+fuel) multiplied by its acceleration, $g$. 
The thrust of the rocket is obtained from the exhaust mass-flow rate times the exhaust velocity. For a non-relativistic case, the needed power, $P$, of the engine thus equals the mass-flow rate times one-half the square of the exhaust velocity. From that we get, $P = M_i\: g u/2$.
For a relativistic exhaust speed ($u \sim c$) it becomes $P = M_i\: g c$.

If at the maximum speed so far achieved, which is 16 km s$^{-1}$ for the New Horizons probe to Pluto, we could make a return trip to the Moon in a little more than half a day (ignoring the slowing down due to the Earth's gravity), a similar return trip at this speed to Proxima Centauri, the star nearest to our solar system, will take about 160,000 years which is over 6,000 human generations, and this is of the order of time that has passed since the homo sapiens (humans) first appeared on the scene. One can thus conclude that in order to reach these interstellar destinations, one would have to travel much faster, in fact with speeds close to that of light, c, which is the maximum attainable speed for any object. Otherwise such a trip would be unimaginable. And to get close to c, we need alternative fuels. 

\section{Various rocket concepts}
\subsection{Chemical fuel rocket}
Till now the chemical energy being used comes from a mixture of liquid oxygen and hydrogen, which yields 100 MJ (Mega Joules) per kg of fuel. The highest efficiency is achieved if the end products of the chemical reactions themselves can be expelled for propulsion with the energy produced. Then one will get an exhaust speed of $u=14$ km/s. Attaining a modest maximum final value of one thousandth of the speed of light, would mean $\sim 17,000$ years of travel time for a return trip to Proxima Centauri, at a distance of 4.24 light years. This would itself require, due to the four stages of the journey, an extremely high mass ratio (${\cal R} \sim (1.001)^{4c/u} \sim 1.6 \times 10^{37}$). This implies that a ten ton payload (a minimum from any standards) will need a fuel $\sim 1.6 \times 10^{44}$) gm, the mass equivalent of $\sim 100$ billion suns or a whole galaxy. Not at all a viable possibility, considered from any angle. Perhaps nuclear fuel might be a better option.

\subsection{Nuclear fuel - fission or fusion?} 
Uranium yields about $6.5 \times 10^7$ MJ/kg of energy through fission, or about a million times better than the chemical reactions. In this case, we could get an exhaust velocity, 12,000 km/s or $u=c/25$, and we could possibly attain a maximum travel speed, $v=0.1c$, which implies ${\cal R}\sim 12$ . Considering, however, four stages of the journey, ${\cal R} > 20,000$ will be needed. A round trip to the nearest star would, however, require a minimum of 170 years of travel time. Relativistic effects of time dilation would be insignificant at such speeds. 

Fusion could provide ten times more energy per unit fuel mass. Despite the fact that controlled reactions of fusion of lighter nuclei have not yet been very successful, we can imagine that the technology required for it could be developed in the years to come. Banking on this assumption, one could propose the energy required for interstellar travel to come from nuclear fusion.

Using fusion  of lighter nuclei, an exhaust speed of c/8.4 may become possible (see Appendix C), and that we could attain a top speed of 0.3c, requiring for a return journey a mass ratio more than 32,000. At these speeds a trip to the nearest star would require for the return journey a minimum of 60 years of total travel time, slightly more than the average working life span of a single generation. Of course a ten ton payload will mean more than 320,000 tons of hydrogen to be carried aboard and to be converted into helium and propelled behind during the journey. This will be $\sim 2 \times10^{17}$ MJ of energy, which is around 400 years worth of total energy consumption ($5 \times 10^{14}$ MJ for the year 2018) of the whole world!

The examples discussed so far were for accelerations much lower than $g$, 
the acceleration due to gravity on the Earth,  an ideal value for journeys made by humans for long durations. In fact, an acceleration of $1g$ could make it possible to attain much higher speeds for the spaceship and thus substantially cut down the travel time. However, as we will show later, the mass ratio, ${\cal R}$, then snowballs to extremely high values, making even the nuclear fusion energy as a mode of locomotion for journey to other stars, not very promising. Thus a vision of interstellar space travel will be highly unrealistic, if we were to depend only on these energy sources.

\subsection{Antimatter rockets}
An antimatter rocket would have a far higher energy density and specific impulse, i.e. total impulse (or change in momentum) delivered per unit of propellant mass, than any other proposed class of rocket. When matter and anti-matter is made to fuse, the entire mass gets converted to radiation, but the technology supporting such a mode of energy production, would require matter and anti-matter to be stored at a safe distance from each other and to be able to combine them, a proper amount, at a proper time in order to be able to use the energy which is produced due to annihilation. 

The problem, however, is that all of the current methods of manufacturing antimatter require enormous particle accelerators and produce antimatter in very small quantities, and to store antimatter, if we need a ton of magnets for one gram of antimatter, the entire idea of a lightweight way to store and carry immense amounts of energy remains no longer meaningful. Antimatter could nevertheless perhaps find use in interstellar spaceships as a way to help trigger nuclear reactions.

\section{Non-rocket concepts}
\subsection{A scoop on the way} 
In a fusion rocket 
a huge scoop could 
collect diffuse hydrogen from the interstellar space
and burn it on flight, 
using proton-proton fusion reaction and expel the fusion product to 
get the thrust. The idea is attractive as the fuel would be collected en route, 
but all attempts to design some kind of a scoop has the unfortunate effect of producing 
more drag than you get back thrust. 

\subsection{Sailing away}
Solar sails are a form of spacecraft propulsion using the solar pressure, of a combination of photons and solar wind from the Sun, to push large ultra-thin mirrors to high speeds. Comets tails are pushed away from the Sun by the same mechanism. 

The momentum of a photon or an entire flux is given by 
$p = E/c$, where 
$E$ is the photon or flux energy, 
$p$ is the momentum. At 1 au the flux density of solar radiation is 
1.36 kW/m$^2$, resulting in a pressure of  
$\sim 4.5 \mu$Pa. A perfectly reflecting sail with 1-sq. km area could thus yield a force 
$\sim 9$ N, while the Sun's gravitational force on one ton mass there is about $6$ N. As both the radiation pressure and the gravity fall with the square of distance from the Sun, a 1-ton load attached to a sail of 1-sq. km area could get pushed outward by the radiation pressure and thus escape the solar system. 

Solar wind on the other hand exerts only a nominal dynamic pressure of about 3 to 4 nPa, three orders of magnitude less than solar radiation pressure on a reflective sail, and would not relatively have much effect. 

A physically realistic approach would be to use the light from the Sun to accelerate. The ship would begin its trip away from the system using the light from the Sun to keep accelerating. Beyond some distance, the ship would no longer receive enough light to accelerate it significantly, but would maintain its course due to inertia. When nearing the target star, the ship could turn its sails toward it and begin to decelerate. Additional forward and reverse thrust could be achieved with more conventional means of propulsion such as rockets.
\subsection{Laser sails or particle beams}
Laser sails might be another way to go. Instead of relying just on the enormous amount of light given off by the Sun, laser sails to Proxima Centauri could also ride laser beams that the  earthlings would fire carefully at those ships to give an extra boost, especially when sails were too far away to catch much light from our Sun. The problem with laser sails is that a lot of light needs to be used for a long time to get fast enough to get to Proxima Centauri within a human lifetime. This means very powerful and extraordinarily large lasers are needed in order to focus on sails that get farther and farther away.

An idea similar to light sails could be firing a particle beam at a spaceship that would ride that energy. The problem with laser beams is that they disperse over distance, so we could use particle beams. The beam would have to have a neutral electrical charge so as not to disperse itself over time.

\subsection{Bombs!}
Another idea for space travel would involve riding explosions through space. Such "pulsed propulsion" would hurl bombs behind a ship, which is shielded with a giant plate. The explosions would push against the plate, propelling the ship.  Nuclear pulsed propulsion works best for really big systems. If we want to send a colony of 1,000 people to space, this might be the way to do it

\section{Some other fanciful ideas}

\subsection{Interstellar travel by transmission}
If physical entities could be decomposed as ``information'',
then transmitted and then reconstructed at a destination, travel at nearly the speed of light would be possible, which for the ``travellers'' would be instantaneous. However, sending an atom-by-atom description of (say) a human body would be a daunting task. Extracting and sending only a computer brain simulation is a significant part of that problem. ``Journey'' time would be the light-travel time plus the time needed to encode, send and reconstruct the whole transmission.

\subsection{Generation-ships}
A generation-ship is a kind of interstellar ark in which crew that arrive at the destination are descendants of those who started the journey. Generation ships are not currently feasible, because of the difficulty of constructing a ship of the enormous required scale, and the great biological and sociological problems that life aboard such a ship raises.

\subsection{Suspended animation}
Scientists and writers have postulated various techniques for suspended animation. These include human hibernation and cryonic preservation. While neither is currently practical, they offer the possibility of sleeper ships in which the passengers lie inert for the long years of the voyage, hopefully without many after-effects.

\section{Other difficulties of interstellar travel}
\subsection{Ex-communication!}
The round-trip delay time is the minimum time taken for to-and-fro communication between the probe and the Earth. For Proxima Centauri this time would be 8.5 years. Of course, in the case of a manned flight the crew can respond immediately to their emergencies. However, the round-trip delay time makes them not only extremely distant from but, in terms of communication, also extremely isolated from the Earth. In fact the communication issue could become the biggest problem. How will the people born in an interstellar colony identify themselves with no attachment to the Earth? Will they not feel literally excommunicated from the Earth?

\subsection{Hard-hitting interstellar medium} 
A major issue with traveling at extremely high speeds is that interstellar dust and gas may cause considerable damage to the craft, due to the high relative speeds and large kinetic energies involved. A robust shielding method to mitigate this problem would be needed.  Larger objects (such as macroscopic dust grains) are far less common, but would be much more destructive. The risks of impacting such objects, and methods of mitigating these risks, will have to be adequately addressed.

\subsection{Manned missions} 
The mass of any craft capable of carrying humans would inevitably be substantially larger than that necessary for an unmanned interstellar probe. 
The requirements for food, water, medical and other life-sustaining needs of the crew will literally put huge burden on the mission. 
In the case of interstellar missions, given the vastly greater travel times involved, there will thus be the necessity of a closed-cycle life support system, which would last over decades. 
In generation ships, will there be a large enough gene pool for healthy future 
generations? 
There will be the ethical 
questions -- Should a new-born be condemned 
to a life-time of journey in which he or she may have no choice whatsoever. Then there is the possibility that the new generations aboard might change their mind and abandon the mission or go elsewhere, keeping no contact with 
the Earth.

\section{A hypothetical journey!}
Let us make a hypothetical journey to Proxima Centauri, the star closest to the Solar system, at a distance of 4.24 light years. For this we expand on a scenario created by Purcell \cite{5},
with the crew always under an acceleration of $1g$, the acceleration due to gravity, so that the they ``feel at home''. From Equation (4) we find that the return trip will take a total of 12 years of the earth time, with the top speed (Equation (2)) reaching $0.95c$ midway point of the journey. However, from Equation (3), the traveller would age only by about 7 years. We already saw that a chemical fuel cannot provide enough thrust as it does not give rise to large enough exhaust speed. So let us try nuclear fusion of hydrogen into helium, for which the best possible exhaust speed is $u=c/8.4$ (see Appendix C). Then assuming a 100\% efficiency, the relativistic rocket equation (5) yields a mass ratio ${\cal R} \sim 4.7 \times 10^6$ to reach a maximum speed $0.95c$. However, if we consider the deceleration and the return journey as well, the scenario becomes impossible as the mass ratio for the nuclear fusion case swells to ${\cal R} \sim (4.7 \times 10^6)^4 \approx 5 \times 10^{26}$. So for a 10 ton payload, we will need a fuel mass of $\sim 5 \times 10^{33}$ gm, that is, equivalent to more than two suns.
Thus one will have to tug along fuel mass equivalent to two suns or more, in order to accomplish a return trip to the nearest star beyond the Solar system. The fuel requirement could be reduced substantially if we are able to somehow achieve nuclear fusion of hydrogen into iron, the ultimate stage in the nuclear fusion, where the maimum exhaust speed becomes $u=c/7.4$ (Appendix C). In that case the fuel needed for a return journey to Proxima Centauri, with a 10 ton payload, reduces to $\sim 3 \times 10^{30}$, equivalent to the mass of $\sim 500$ earths. Still an impossible amount of fuel.

Though recently an earth-size planet has been found orbiting around $\alpha$-Centauri B, but it seems too close to the parent star and would be very hot and perhaps not habitable. It is estimated that to visit a habitable planet and hopefully encounter some extraterrestrial life,  we may have to probe stars up to about 12 light years. For instance, Ross 128 b, a confirmed Earth-sized exoplanet, orbiting within the inner habitable zone of the red dwarf Ross 128, lies at a distance of about 11 light years from the Earth. Another exoplanet, Luyten b, orbiting within the habitable zone of the red dwarf Luyten's Star, is at a distance of 12.2 light years from our Solar system. With this in mind, let us make a hypothetical return trip to an exoplanet, say, at a distance of 12 light years. From Equation (4) we find that the return trip will take a total of 28 years of the earth time, with the top speed (Equation (2)) reaching $0.99c$ midway point of the journey. However, from Equation (3), the traveller would age only by about 10 years. For  the best possible exhaust speed is $u=c/7.4$, to reach $0.99c$, the mass ratio for the nuclear fusion case swells to ${\cal R} \sim 2 \times 10^{34}$. So for a 10 ton payload we will need a fuel mass of $\sim 2 \times 10^{41}$ gm, that is, equivalent to $\sim 100$ million suns. This would imply consuming, throughout the journey of 10 years on board, on the average, fuel mass about one third of the sun every second. This means the energy that the Sun produces during its life time of $\sim 10^{10}$ years, would be consumed every three seconds to accelerate the spaceship. In fact the fuel consumption will be orders of magnitude higher in the initial stages, being at a rate $M_i\: g/u$, that is $\sim 25$ suns per second. A scenario not imaginable even in the wildest of our fantasies. 

Thus forgetting the chemical fuel, even the nuclear fusion could not be the source of energy for interstellar travel. And that too when we restricted travels to only a few light years within the reach of the Solar System. It is quite clear that, one would need an exhaust speed, $u \approx c$ and matter-antimatter annihilation only may provide it. To reach $0.99c$, the mass ratio in such a case may appear to be manageable, ${\cal R} = 14$, at least for one leg of the journey, which however, snowballs to ${\cal R} = (14)^4 = 40,000$ for the complete journey in four stages, implying 200,000 tons of matter and antimatter each. For the early part of the journey we will need $\sim 1.2\times 10^{12}$ MW, about seven times more than the radiation that the Earth receives from the Sun. But with all that in gamma-rays, our problem will be not only to shield the payload but also to shield the Earth. Again, not a very promising scenario! 

\section{Could we? Or should we?}
So far no one has created technology that is widely agreed upon as capable of caring for or preserving humans across the lifetimes it might take to get to even Proxima Centauri; it might easily take more than one lifetime to reach any star system! If that is so, mission designers might have to take procreation and family into account so that offspring of the original crew would get properly educated and trained to manage the ship in due course.

Thus a trip to our nearest star requires not only ingenious methods of propulsion and a minimum of decades en route, but also a sophisticated system of life support for the human crew to survive the journey. Not only the costs and difficulties are almost insurmountable, but they would also require almost unparalleled public and governmental support. The ultimate question then might change from -- Could we to should we?

Even if the constraints imposed by the technology are ignored, the requirement of energy plays a huge constraint by itself. A huge amount of fuel would have to be put to use for such an endeavour and many generations of earthlings would have to work on such a project.

There is a very strong likelihood that the mission would fail due to many other factors. We have ignored the requisites of food and water and other medicinal requirements for the crew. There is also the effect of the harmful radiation such as cosmic rays and impacts with other larger bodies. What if some deadly disease strikes? It is unlikely that living beings will be able to survive such ordeals for time periods of the order of decades.

Further we have not even considered the time and resources needed for possible research and conduction of experiments at the place of the destination, without which such a trip would not be of much advantage to us, anyway.

\section{Conclusions} 

Taking these severe limitations into account, we can conclude that space travel, even in the most distant future, will remain confined to our own planetary system, and a similar conclusion will hold forth for any other civilization, no matter how advanced it might be, unless those extraterrestrial species have life spans order of magnitude longer than ours. 
Even in such a case it is unlikely that they will travel much farther than their immediate stellar neighbourhood, as each such excursion will exhaust the resources of their home planet so much that those will dwindle rather fast and there might not be much left for the further scientific and technological advancements. 
So the science-fiction fancy of a Galactic Empire may ever remain in our fantasies only. 
And as for the mythical UFOs, whose quiet appearances do get reported in the press once in a while, recent explorations have shown no evidence that any such thing could have an origination within our own solar system itself, a ``quiet'' return trip from a distant star is almost impossible as it could not be so quiet as the exhaust in any such trip will dazzle the sky like many suns or perhaps more like a gamma ray burst occurring, but not in a distant part of the universe, instead going off right in our own solar backyard.
%--------------------------------------------------------------------
\section*{Appendix A: The distance-time relation for an accelerated motion with relativistic speeds} 
We can compute time $T$ of a spaceship traveller, undergoing a proper acceleration $g$ to achieve relativistic speeds, in terms of the time $t$ and distance $x$, as measured by a set of observers stationary with respect to the launching station. 
We assume it to be a 1-dimensional motion, say, along the $x$-axis, taking $x=0$ and $t=0$  at the start of the journey at $T=0$.  From relativistic transformations, we have the time dilation formula,  
${\rm d} t=\gamma\: {\rm d}T$ while for the longitudinal acceleration we have, ${\rm d}{v}/ {\rm d} t={g}\:\gamma^{-3}  $ \cite{1}.  

The equation of motion then is
\begin{eqnarray*}
\gamma^{3}{\rm d}{v} = g\: {\rm d} t=g \gamma \:{\rm d}T .
\end{eqnarray*}
We can integrate it 
\begin{eqnarray*}
\int_{0}^{v} \frac {{\rm d} v}{1-(v/c)^2} = \int_{0}^{T} g \: {\rm d}T .
\end{eqnarray*}
For a constant proper acceleration $g$,  we thus get
\begin{eqnarray*}
\frac{v}{c}= \tanh (gT/c),
\end{eqnarray*}
which gives $\gamma=\cosh (gT/c)$. 

From this we can get a relation between $t$ and $T$ as
\begin{eqnarray*}
\int_{0}^{t} {\rm d}t = \int_{0}^{T} \gamma\: {\rm d}T = \int_{0}^{T} \cosh \frac{gT}{c} \: {\rm d}T,
\end{eqnarray*}
or
\begin{eqnarray*}
t = \frac{c}{g} \sinh \frac{gT}{c}. 
\end{eqnarray*}
The distance covered is
\begin{eqnarray*}
\int_{0}^{x} {\rm d}x =\int_{0}^{t} v \:{\rm d}t = c \int_{0}^{T} \sinh \frac{gT}{c} \: {\rm d}T,
\end{eqnarray*}
or
\begin{eqnarray*}
x = \frac{c^2}{g} \left[\cosh \frac{gT}{c} -1\right].
\end{eqnarray*}
Distance $x$ can be expressed in terms of $t$ as
\begin{eqnarray*}
x = \frac{c^2}{g} \left[\sqrt{1+\left(\frac{gt}{c}\right)^2} -1\right].
\end{eqnarray*}
\section*{Appendix B: The relativistic rocket equation}
If in the instantaneous rest frame of the rocket, a fuel mass $\Delta m$ is consumed during a proper time $\Delta T$, to generate energy that causes the expulsion of the propellent with an exhaust speed $u$, with a corresponding Lorentz factor $\gamma_u= 1/\sqrt{1-(u/c)^2}$, from the  energy conservation we have, $\gamma_u\Delta m' c^2 = \Delta m c^2$, where $\Delta m'$ is the mass in the expelled fuel's rest frame. The expelled mass carries a momentum, $\gamma_u\Delta m' u= \Delta m u$ and from momentum conservation,  we get  
%caused by the exhaust fuel propels the spaceship with a proper acceleration $g$, = \Delta m u = M g$, which gives
\begin{equation*}
 M g=-\frac{{\rm d}M}{{\rm d}T} u , 
\end{equation*}
Using ${\rm d}T= {\rm d}t/\gamma$ and ${g}=\gamma^{3}\:{\rm d}{v}/ {\rm d} t$ from Appendix A, we get
\begin{equation*}
 M \gamma^2 \frac{{\rm d} v}{{\rm d}t}=-\frac{{\rm d}M}{{\rm d}t} u , 
\end{equation*}
  or
\begin{equation*}
 \frac{{\rm d} v}{1-(v/c)^2}=-\frac{{\rm d}M}{M}u .
\end{equation*}
We can integrate it
\begin{equation*}
\int_{0}^{v}\frac{{\rm d} v}{1-(v/c)^2}=-u\int_{M_i}^{M_f}\frac{{\rm d}M}{M},
\end{equation*}
 to get 
 \begin{eqnarray*}
 \tanh^{-1} \frac{v}{c}=\frac{u}{c} \ln {\cal R},
\end{eqnarray*}
which can be written as
\begin{eqnarray*}
\frac{v}{c} = \tanh [\ln {\cal R}^{{u}/{c}}]= \frac{{\cal R}^{u/c} - {\cal R}^{-u/c}} {{\cal R}^{u/c} + {\cal R}^{-u/c}}.
\end{eqnarray*}
The relativistic rocket equation then is 
\begin{equation*}
\frac{v}{c} = \frac{1 - {\cal R}^{-2u/c}} {1 + {\cal R}^{-2u/c}}, 
\end{equation*}
or
\begin{equation*}
{\cal R}  = \left[\frac{(1 + \frac{v}{c})}{(1 - \frac{v}{c})}\right]^{c/2u}=\left[\gamma{\left(1 + \frac{v}{c}\right)}\right]^{c/u}.
\end{equation*}
For a constant proper acceleration $g$,  we substitute for $v$ and $\gamma$ from Appendix A, to get
\begin{eqnarray*}
{\cal R}  &=& [\cosh (gT/c) + \sinh (gT/c)]^{c/u}\\
&=&\exp(gT/u).
\end{eqnarray*}
\section*{Appendix C: The exhaust velocity limit for a nuclear fusion rocket}
In a nuclear fusion reaction of hydrogen into helium, an amount $\epsilon=0.71\%$ of the fuel mass gets converted into energy, while for a conversion from hydrogen to iron, the ultimate stage in the nuclear fusion, the amount is $\epsilon=0.92\%$ \cite{6}. 

The energy released by this amount could be converted into the kinetic energy [$(\gamma_u-1)\Delta m' c^2$] of the expelled fuel mass, giving 
\begin{eqnarray*}
(\gamma_u-1)\Delta m' c^2=\epsilon \Delta m c^2.
\end{eqnarray*}
Using $\gamma_u\Delta m'= \Delta m$ (Appendix B), we get
\begin{eqnarray*}
(1-\sqrt{1-(u/c)^2})=\epsilon .
\end{eqnarray*}
This yields for a nuclear fusion rocket, the best possible values for the exhaust speed, $u=c/8.4\:,\:$ for $\epsilon=0.71\%$\\  and\\  $u=c/7.4\:,\:$ for $\epsilon=0.92\%$.\\\\


\begin{thebibliography}{20}
\bibitem{5} Purcell, E., Interstellar Communication, ed. Cameron, A. G. W.,  Benjamin Inc. (1963), p. 121-143.
%\bibitem{1} Andrews, B., Astronomy, July 2012, p. 22-27.
\bibitem{2} Sagan, C., Carl Sagan's Cosmic Connection - An Extraterrestrial Perspective, Cambridge University Press (2000).
\bibitem{3} Sagan, C., Pale Blue Dot: A Vision of the Human Future in Space, Ballantine Books (1997).
%\bibitem{4} Taylor, E. F. \& Wheeler, J. A., Space Time Physics, 2nd ed., Freeman, New York (1992). 
\bibitem{1} Jackson, J.D.: Classical Electrodynamics, 2nd edn., Wiley, New York (1975).
%\bibitem{71} Tolman R C 1934 {\it Relativity thermodynamics and cosmology} (Clarendon: Oxford).
\bibitem{6} von Hoerner, S., Interstellar Communication, ed. Cameron, A. G. W.,  Benjamin Inc. (1963), p. 144-159.
\end{thebibliography}
\end{document}